\documentclass[11pt]{article}
\usepackage{moriond,amsmath,amssymb,scalefnt}

\def\Journal#1#2#3#4{{#1} {\bf #2}, #3 (#4)}

\def\PLB{{\em Phys. Lett.}  B}
\def\PRL{\em Phys. Rev. Lett.}
\def\PRD{{\em Phys. Rev.} D}

\def\JHEP{\em JHEP} 
\def\CPC{\em Comput. Phys. Commun.}
\def\JCP{\em J. Comput. Phys.}

\def\be{\begin{equation}}
\def\ee{\end{equation}}
\def\bea{\begin{eqnarray}}
\def\eea{\end{eqnarray}}

\def\eqs#1#2{{Eqs.~(\ref{#1})--(\ref{#2})}}
\newcommand{\abbrev}{\scalefont{.9}}
\newcommand{\msbar}{$\overline{\mbox{\abbrev MS}}$}

\newcommand{\Photo}{}

\begin{document}
\vspace*{4cm}
\title{THE COST OF GAUGE COUPLING UNIFICATION IN THE   $SU(5)$ MODEL AT
  THREE LOOPS}

\author{ L. MIHAILA }

\address{Institut f\"{u}r Theoretische Teilchenphysik,
Karlsruhe Institute of Technology (KIT),\\ D-76128 Karlsruhe, Germany}

\maketitle\abstracts{
The non-supersymmetric $SU(5)$ model can accommodate heavy neutrinos and
gauge coupling unification when augmented with an adjoint fermionic
multiplet $24_F$. Among the most important phenomenological implications
of the model is the prediction of light fermions and scalars, charged
under the $SU(2)_L$ gauge group, in the reach of the Large Hadron
Collider (LHC). In this talk, we report on the recent calculation~\cite{DiLuzio} of the
correlation function between the mass scale of the new electroweak  multiplets and the
gauge coupling unification scale at three loop accuracy.}

\section{Introduction}
Nowadays, it is well established that the minimal non-supersymmetric
$SU(5)$ model in its original form~\cite{Georgi:1974sy} is
phenomenologically ruled out. In principle, this is due to the lack of gauge coupling
unification and the massless neutrinos.  These unsatisfactory
aspects of the original model can be simultaneously eliminated, if one adds  
an additional fermionic multiplet in the adjoint representation
$24_F$~\cite{Bajc:2006ia,Bajc:2007zf}. 
To understand the role of the $24_F$ multiplet in the model, let us
recall its decomposition w.r.t.   the Standard Model (SM) gauge group
$SU(3)_c\times SU(2)_L\times U(1)_Y$.  
\begin{eqnarray}
24_F &=& \underbrace{(1,1,0)_F}_{S_F} \oplus \underbrace{(1,3,0)_F}_{T_F}
\oplus \underbrace{(8,1,0)_F}_{O_F} \oplus 
\underbrace{(3,2,-\tfrac{5}{6})_F}_{X_F} \oplus
\underbrace{(\overline{3},2,+\tfrac{5}{6})_F}_{\overline{X}_F}  
\, ,
\end{eqnarray}
where $S_F$, $T_F$, $O_F$ ($X_F$) are Majorana (Dirac) degrees of
freedom. 
 A special role in the model is played by the electroweak singlet and triplet states 
$S_F$ and $T_F$. They are involved in the Yukawa interactions that
after the SU(5) gauge symmetry breaking will generate masses for
neutrinos through a hybrid type-I+III seesaw mechanism
\cite{Minkowski:1977sc,GellMann:1980vs,Yanagida:1980xy,Mohapatra:1979ia,Schechter:1980gr}.
The electroweak singlet $S_F$ resembles a sterile neutrino, whereas the electroweak 
triplet is sometimes referred to as a heavy lepton.\\
As can be read from the above decomposition, the fermionic states $T_F,
O_F , X_F$ are charged under the groups $SU(2)_L$, $SU(3)_C$ and $SU(3)_c\times
SU(2)_L\times U(1)_Y$, respectively. Thus, they will give also
contributions to the gauge coupling evolution with the energy scale. In
particular,  the electroweak triplets $T_F$ can  delay the meeting of the couplings
$\alpha_1$ and $\alpha_2$ from the energy scale of about $10^{13}$~GeV in the minimal
$SU(5)$ model to values in agreement with the bounds enforced by the
non-observation of proton decay~\cite{Nishino:2012ipa} of about
$10^{15.5}$~GeV. The underlaying condition is that $T_F$ states  are rather light,
in the TeV range, in order to  have the maximal impact on the evolution
of $\alpha_2$. In contrast, the states $X_F$, that are charged both
under the $SU(2)_L$ and $U(1)_Y$ have always  the opposite effects, due
to their  contributions to the beta functions of the coupling constants
$\alpha_1$ and $\alpha_2$. Thus, in order to reach a high enough
unification scale for the coupling $\alpha_1$ and $\alpha_2$, one needs
in addition a very heavy mass scale for $X_F$ states. However, this mass
scale can be at most of the order of $M_G^2/\Lambda$, where $M_G$
denotes the unification scale and $\Lambda$ is the cutoff scale of
the $SU(5)$ model. The latter have to be chosen in such a way that the
low-energy value of the ratio $m_b/m_{\tau}$ (where $m_b$ stands for the
bottom quark mass and $m_\tau$ for the tau lepton mass) is correctly
reproduced in the model and to maximize the perturbativity domain.  It
was shown~\cite{Bajc:2006ia,Bajc:2007zf} that a value of about
$\Lambda=100\, M_G$ is a  reasonable choice. \\
Furthermore, for a complete unification it is also necessary that the
strong coupling constant $\alpha_3$ meets the electroweak couplings $\alpha_1$ and $\alpha_2$ at
the right  energy scale. Obviously, the states that have a direct impact
on the energy evolution of $\alpha_3$ are the colour octet fermions
$O_F$. As we show in the next section, a proper unification requires
that the states $O_F$ live at intermediate mass-scale of about $10^8$~GeV.

 The crucial parameter for phenomenology is actually
the effective mass of the electroweak triplet states. This mass scale is
defined as an average between the mass scale of the   electroweak
triplet fermions $T_F$ and the similar components of the scalar
multiplet that lives in the
$24$-dimensional representation~\footnote{We denote the latter states by $T_H$ to
distinguish their scalar origin.}.    Both types of  triplets, if light
enough, can give interesting signature 
at the LHC.   
The fermionic components  lead to lepton number violation effects in
same sign di-lepton events~\cite{Bajc:2006ia,Bajc:2007zf}. The bosonic
triplet instead can easily modify  
the decay properties of the Higgs boson (see e.g.~\cite{Chang:2012ta}), 
that will be measured with increasing precision at the LHC. 

Let us also mention  that,  the Higgs sector is the one of the genuine
SU(5) model 
\begin{eqnarray}
5_H &=& \underbrace{(3,1,-\tfrac{1}{3})_H}_{\mathcal{T}} \oplus \underbrace{(1,2,+\tfrac{1}{2})_H}_h
\quad \mbox{and}\\
%\ee
%and
%\be
24_H  &=& \underbrace{(1,1,0)_H}_{S_H} \oplus \underbrace{(1,3,0)_H}_{T_H}
\oplus \underbrace{(8,1,0)_H}_{O_H}  \oplus   
\underbrace{(3,2,-\tfrac{5}{6})_H}_{X_H} \oplus
\underbrace{(\overline{3},2,+\tfrac{5}{6})_H}_{\overline{X}_H}  
\, ,
\end{eqnarray}
where $S_H$, $T_H$ and $O_H$ ($\mathcal{T}$, $h$ and $X_H$) are real (complex) scalars. 
In our notation, $h$ stands for the SM Higgs doublet.  The 
mass spectrum of the model is derived as usual from the  minimization
conditions of the scalar potential.
 In this respect,   
it is  a nontrivial fact that the tree-level calculation of the spectrum 
allows the mass pattern required by unification
\be
m_{T_F}\approx m_{T_H} \ll m_{O_F}\approx m_{O_H}
\ll M_G\,.
\label{eq:unifspect}
\ee
Nevertheless,  
the required mass hierarchy strengthen the fine-tuning issue typical for
non-supersymmetric GUTs. 

\section{Framework}
The effective mass scale for the electroweak triplets can be determined
from the constraint of gauge coupling unification. More precisely, it
 depends only on the unification scale of the electroweak
couplings $\alpha_1$ and $\alpha_2$ at one- and two-loop oder in
perturbation theory. The dependence on the strong coupling constant
occurs starting from three loops. To study the energy evolution of the
electroweak couplings  for the case of a largely split mass spectrum
 as required in Eq.~(\ref{eq:unifspect}),  it is convenient to
apply the method of effective field theories (EFT)s.
It consists in integrating out the heavy degrees of freedom that 
cannot influence the physics at the low-energy scale. \\
In physical renormalizations schemes like the momentum subtraction
scheme or the on-shell scheme, the  effects due to heavy particle
thresholds are encountered in
the renormalization constants of the parameters.  However, for the analysis of the gauge coupling
unification that requires the running of the couplings over many orders
of magnitude, higher order radiative corrections to the RGEs are
essential. But, their calculation beyond 
one-loop order in mass dependent 
renormalization schemes is quite involved. A much more suited scheme for
this purpose is the minimal subtraction scheme
(\msbar), for which 
the gauge coupling beta functions are mass independent and their computation is
substantially simplified. Nevertheless, in this scheme the
Appelquist-Carazzone~\cite{Appelquist:1974tg}  
theorem does not hold anymore and  
the threshold effects have to be taken into account explicitly. The
latter are parametrized through the 
decoupling ( matching ) coefficients. They can be computed 
perturbatively using the physical constraint that the Green's functions involving light particles
have to be equal in the original and the effective theory. For the
computation presented here, we adopt this second method 
and apply it up to the  
third order in perturbation theory.

The computation of the renormalization
constants up to the three-loop order in the \msbar{} scheme can be reduced to the evaluation of
only massless propagator  diagrams. For the present calculation, we use a well-tested chain of programs: 
the Feynman rules of the model are obtained with the help of the program
 {\tt FeynRules}~\cite{Christensen:2008py} and translated into 
 {\tt QGRAF}~\cite{Nogueira:1991ex}  syntax. {\tt QGRAF} generates further all contributing Feynman
 diagrams. The 
output is passed via {\tt
  q2e}~\cite{Harlander:1997zb,Seidensticker:1999bb}, which transforms
Feynman diagrams into 
Feynman amplitudes, to {\tt exp}~\cite{Harlander:1997zb,Seidensticker:1999bb} 
that generates {\tt FORM}~\cite{Vermaseren:2000nd} code. The latter is
processed by {\tt   MINCER}~\cite{Larin:1991fz} 
that computes analytically massless propagator diagrams up to three
loops  and outputs the  $\epsilon$ expansion of the result.
 The three-loop expressions for the beta functions of the gauge couplings in the
low-energy theories can be found in Ref.~\cite{DiLuzio}.

For the computation of the matching coefficients of the gauge couplings,
 when two  different theories  are matched together, one has to
consider Green's functions involving light particles and a vertex that contains
the gauge couplings $\alpha_i$. Since the matching coefficients are
universal quantities, they must be independent of the momentum transfer
of the specific process taken under consideration. For convenience of
the calculation, one chooses vanishing external momenta. 
Thus, in dimensional regularization only diagrams containing at least
one heavy particle inside the loops contribute  have to be taken
into account. As a consequence,  the resulting   Feynman amplitudes can be
 mapped to massive tadpole topologies that are handled with
the help of the program MATAD~\cite{Steinhauser:2000ry}. Explicit
two-loop results for the matching coefficients of the gauge couplings  in
the $SU(5)+24_F$ model can be found in Ref.~\cite{DiLuzio}.

\section{Numerical Results}
%%%%%%%%%%%%%%%%%%%%%%%%%%%%%%%%%%%%%%%%%%%%%%%%%%%%%%%%%%%%%%
In this section we study the numerical impact of the three-loop  corrections 
on the evolution of the gauge couplings and on the correlation function
between the electroweak triplet mass and the GUT scale. In practice,
 we integrate numerically the $n$-loop beta functions of the 
gauge couplings taking into account also the $(n-1)$-loop 
running of the top-Yukawa 
coupling and the $(n-2)$-loop running of the Higgs boson self-coupling
together with the $(n-1)$-loop matching conditions for the gauge
couplings. In the present analysis $n=1,2,3$.
 We can safely neglect the contribution of the bottom and tau Yukawa
couplings. We also neglect in this analysis   
the effects due to the new scalar self-interactions 
of the scalar triplet $T_H$.  
As input parameters for the running analysis we take \cite{Mihaila:2012pz}
\begin{eqnarray}
%\label{alpha1MZ}
\alpha_1^{\overline{\text{MS}}}  (M_Z) &=& 0.0169225 \pm 0.0000039 \, ,  \\
%\label{alpha2MZ}
\alpha_2^{\overline{\text{MS}}}  (M_Z) &=& 0.033735 \pm 0.000020 \, ,  \\
%\label{alpha3MZ}
\alpha_3^{\overline{\text{MS}}}  (M_Z) &=& 0.1173 \pm 0.00069 \, ,  \\
%\label{alphatMZ}
\alpha_t^{\overline{\text{MS}}}  (M_Z) &=& 0.07514 \, ,
\end{eqnarray}
given in the full SM, i.e.~with the top quark threshold effects taken into account. 
 The Higgs self-coupling   is determined assuming a Higgs
 boson with mass $125$ GeV. Thus, we obtain
\be
\label{allamH}
\alpha_{\lambda_h} \approx 0.010 \, . 
\ee

\begin{figure}
\includegraphics[width=0.9\linewidth]{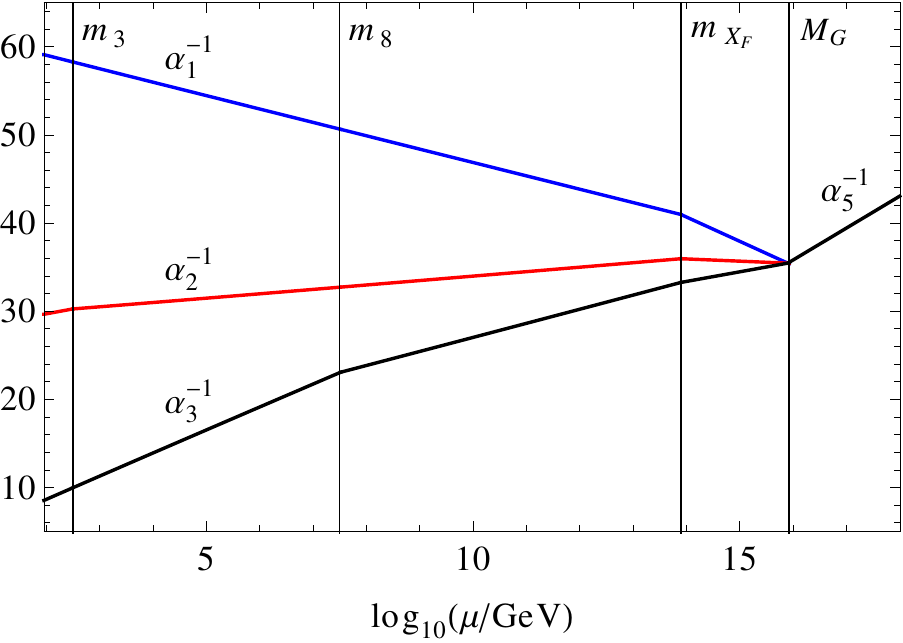}
\caption{\label{sampleunif}
Sample three-loop unification pattern for 
$m_{T_F} = m_{T_H} = 10^{2.5} \, \mbox{GeV}$, 
$m_{O_F} = m_{O_H} = 10^{7.5} \, \mbox{GeV}$,
$m_{X_F} = M_G / 100$ and 
$m_{\mathcal{T}} = m_{X_V} = M_G$. 
The lines with different slopes from top to bottom correspond to 
$\alpha_1^{-1}$ (blue), $\alpha_2^{-1}$ (red) and $\alpha_3^{-1}$ (black). 
The dashed vertical lines denote the masses of the intermediate-scale thresholds. 
}
\end{figure}

For illustration we show in Fig.~\ref{sampleunif} an unification pattern
for the inverse of the gauge couplings, taking into account three-loop
order RGEs and two-loop order threshold corrections. For the intermediate
mass scales we choose: $m_{T_F} = m_{T_H} = 10^{2.5} \, \text{GeV}$, 
$m_{O_F} = m_{O_H} = 10^{7.5} \, \text{GeV}$ and
$m_{X_F} = M_G / 100$. Here, $M_G$ is defined as the scale where the
electroweak couplings $\alpha_1$ and $\alpha_2$ meet. 
 The colour triplet Higgs $\tau$ and the $SU(5)$ gauge bosons $X_V$ 
 also play a role for the unification. However, their effects are
 sub-leading as compared to those generated by the electroweak triplets $T_{F,H}$
 and colour octets $O_{F,H}$, because they are predicted to live at
 very high energies between the unification and the Planck scales.  
 For convenience, we fix their masses at the unification scale
 $m_{\mathcal{T}} = m_{X_V} = M_G$. The vertical lines mark the scales
 where the electroweak triplet $T_{H,F}$ and color octet states
 $O_{H,F}$ are decoupled. At lower order in perturbation theory it is
 advisable to choose the decoupling scale at the effective mass scales
 defined as $m_3=(m_{T_F}^4 m_{T_H})^{1/5}$ and $m_8=(m_{O_F}^4
 m_{O_H})^{1/5}$. 
 The dependence of the physical observables like the value of the
 unification scale or of the unified gauge coupling become more and more
 insensitive on this unphysical parameter, once higher order corrections
 are taken into account.\\
In order to quantify the impact of the newly computed corrections 
let us mention that for such a sample unification pattern 
the relative difference between the two- and three-loop values 
of $\alpha_1$, $\alpha_2$ and $\alpha_3$ evaluated at $M_G$ 
amounts to $0.015 \%$, $0.061 \%$ and $0.08 \%$ respectively. 
This has to be compared with the  relative experimental uncertainties:
$\Delta \alpha_1 / \alpha_1 = 0.023 \%$, $\Delta \alpha_2 / \alpha_2 = 0.059 \%$ 
and $\Delta \alpha_3 / \alpha_3 = 0.59 \%$. 
Hence, for $\alpha_1$ and $\alpha_2$ the three-loop corrections are 
of the same order of magnitude as the experimental uncertainties.

\begin{figure}[ht]
\includegraphics[width=0.8\linewidth]{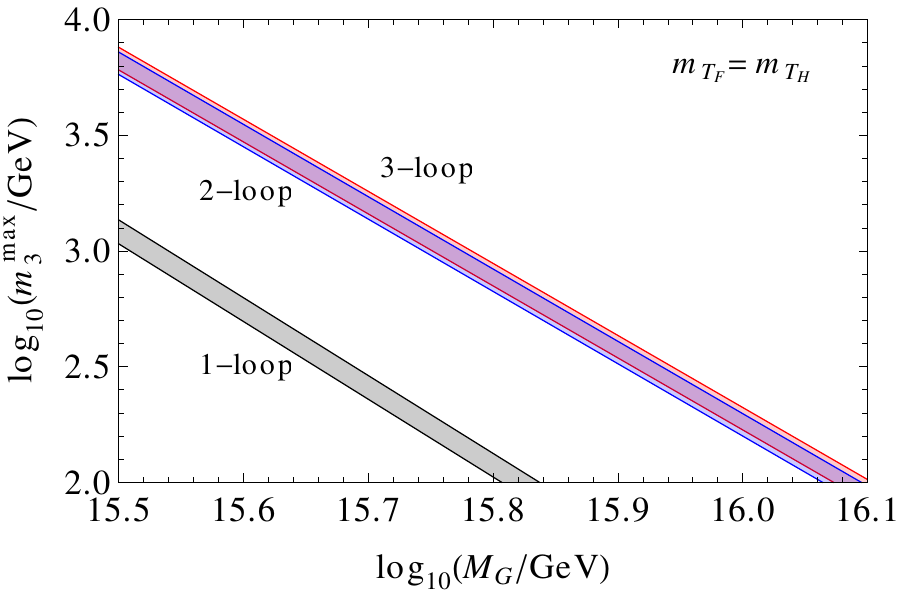}
\caption{\label{MGvsm3} 
$m^{\rm{max}}_3 - M_G$ correlation between. 
The black, blue and red bands (from bottom-left to top-right) correspond respectively to the one-, two- and three-loop 
running analysis. The error bands are obtained by varying the low-energy couplings $\alpha_1 (M_Z)$ 
and $\alpha_2 (M_Z)$ into their $1\sigma$ values (cf.~\eqs{alpha1MZ}{alpha2MZ}).
%$1\sigma$ error bands.
}
\end{figure}

The effective triplet mass scale is an important parametr by
itself. More precisely, its upper bound represents the worse case
scenario for the posibility to observe such  states at
the  LHC. The maximal value of the $m_3$ parameter is  obtained when the
masses of the  
super heavy particles $m_\tau$ and $m_{X_F}$ are set to their maximally
allowed values. Apart from these parameters, $m_3$  depends only on the gauge coupling
unification scale. At the one-loop order, this dependence is known
completely analytically. Starting from two-loop order,  one has to solve a
system of coupled differential equations. Its solution is shown in
Fig.~\ref{MGvsm3} as a function of the unification scale. From the low
left corner to the high right one the corelation function between the
maximal value  $m_3^{\rm max}$ and the
unification scale $M_G$ is shown at one-, two-and three-loop order accuracy. As
can be read from the figure the predictions for $m_3^{\rm max}$ at one-
and two-loop orders  differ by several TeV. In turn, this translates into
a variation of the unification scale by about an oder of magnitude. To
be able to use the electroweak triplet mass scale as a validity check
for the $SU(5)+24_F$ model, we need more precise theoretical
predictions, at least in the range of experimental precision. This
requirement is nicely fulfilled at the three-loop order in perturbation
theory, for which  the theoretical uncertainties are reduced by about a
factor ten.

The upper bound on the electroweak triplet mass scale is actually the parameter
relevant for the phenomenology. It can be used as a validity test of the
model in the sense that if the electroweak triplet states escape
detection at the LHC, i.e. they live beyond the TeV scale, than the
predicted unification scale of the model  
should be below $10^{5.5}$~GeV. This in turn renders the proton lifetime
to be in the reach of the future generation of megaton-scale experiments. Thus, non-observation of
the electroweak triplet states in the TeV range as well as of proton instability are
sufficient to refute the model.\\
In this talk, we present the recent computation  of three-loop order
corrections to the predicted mass scale for the electroweak
tripltes. These higher order corrections are necessary in order to
reduce the theoretical uncertainties on a level compatible with 
those induced by the experimenatal accuracy on the determination of the
electroweak couplings at low energies. Moreover, from a theoretical
point of view, the three-loop order corrections are indispensable in
order to establish the convergence of the perturbative series. This can
be understood from the fact that the relative difference between the
one- and two-loop order corrections amounts to more than 100\%. In
contrast, the three-loop corrections lay on top of the two-loop ones
(see Fig.~\ref{MGvsm3}) and reduce the relative errors at around 25\%,
in the range of the experimental precision.

\section*{Acknowledgments}
The speaker is greatful to Luca Di Luzio for the collaboration on this project. 
This work was supported by the DFG through the SFB/TR 
9 ``Computational Particle Physics''.

%\section*{Appendix}

% We can insert an appendix here and place equations so that they are
%given numbers such as Eq.~\ref{eq:app}.
%\be
%x = y.
%\label{eq:app}
%\ee

\end{document}